\documentclass{aa}  

\usepackage{graphicx}
\usepackage{multirow}
\usepackage{amsmath}
\usepackage{txfonts}
\usepackage{natbib}
\bibpunct{(}{)}{;}{a}{}{,} 
\def\ltsim{\raise 2pt \hbox {$<$} \kern-1.1em \lower 4pt \hbox {$\sim$}}
\def\gtsim{\raise 2pt \hbox {$>$} \kern-1.1em \lower 4pt \hbox {$\sim$}}

\begin{document} 

\title{
New  JVLA observations at 3 GHz and 5.5 GHz \\
of the `Kite' radio source in Abell~2626}

\author{A. Ignesti\inst{1,2}, M. Gitti\inst{1,2}, G. Brunetti\inst{2},
  L. Feretti\inst{2}, G. Giovannini\inst{1,2}}

\institute{
  Dipartimento di Fisica e Astronomia, Universit\`a di Bologna, via Gobetti 93/2, 40129 Bologna, Italy \\
  \email{ myriam.gitti@unibo.it} 
\and 
INAF, Istituto di Radioastronomia di Bologna, via Gobetti 101, 40129 Bologna, Italy }

\authorrunning{Ignesti et al.}
\titlerunning{New  JVLA observations at 3 GHz and 5.5 GHz \\
  of the `Kite' radio source in Abell~2626}

\date{Accepted }

\abstract 
{
  We report on new JVLA observations performed at 3 GHz and 5.5 GHz of
  Abell 2626. The cluster has been the object of several studies in
  the recent years due to its peculiar radio emission, which shows a
  complex system of symmetric radio arcs characterized by a steep
  spectrum. The origin of these radio sources is still unclear. Due to
  their mirror symmetry toward the center, it has been proposed that
  they may be created by pairs of precessing jets powered by the inner
  AGN. }
{
  The new JVLA observations were requested with the specific aim of
  detecting extended emission on frequencies higher than 1.4 GHz, in
  order to constrain the jet-precession model by analyzing the
  spectral index and radiative age patterns alongs the arcs.}
{
  We performed a standard data reduction of the JVLA datasets with the
  software CASA. By combining the new 3 GHz data with the archival 1.4
  GHz VLA dataset we produced a spectral index maps of the extended
  emission, and then we estimated the radiative age of the arcs by
  assuming that the plasma was accelerated in moving hot-spots tracing
  the arcs.}
{
  Thanks to the high sensitivity of the JVLA, we achieve the detection of
  the arcs at 3 GHz and extended emission at 5.5 GHz. We measure a
  mean spectral index $<-2.5$ for the arcs up to 3 GHz. No clear
  spectral index, or radiative age, trend is detected across the arcs
  which may challenge the interpretation based on precession or put
  strong constraints on the jet-precession period. In particular, by
  analyzing the radiative age distribution along the arcs, we were
  able to provide for the first time a time-scale $< 26$ Myr of the
  jet-precession period. } {}

\keywords{
galaxies: clusters: individual: Abell~2626;
galaxies: individual: IC5338, IC5337;
galaxies: jets;
radio continuum: galaxies;
radiation mechanisms: non-thermal;
methods: observational}

\maketitle

\section{Introduction} 

In recent years, radio observations have revealed that a fraction
of galaxy clusters hosts diffuse synchrotron emission on cluster
scale. The discovery of radio sources not associated with any
individual galaxy proves the presence of non-thermal components, such
as relativistic particles and magnetic fields, mixed with the thermal
intra-cluster medium (ICM) on spatial scale that are comparable to the
cluster size. According to their morphology and location in the
cluster, cluster-scale radio sources are classified as radio relic,
radio halos and radio mini-halos \citep[e.g.][]{Feretti_2012}.  Relics
are polarized, elongated, arc-shaped synchrotron sources located in
the peripheries of dynamically disturbed clusters. Radio relics are
also unique probes of the properties of magnetic fields in the
outskirts of galaxy cluster.  Halos and mini-halos are instead 
roundish radio sources located in the cluster central regions. The two
classes differ in size and in the dynamical properties of the hosting
clusters. Radio halos are located in the center of dynamically disturbed
clusters, which show signs of recent major merger activity, whereas
mini-halos are detected only in relaxed, cool-core clusters.  It is
currently thought that relics and halos originate from complex
acceleration mechanisms that are driven by shocks and turbulence in
the ICM \citep[e.g.][]{Brunetti-Jones_2014}.
On the other hand, the mini-halo diffuse emission is always observed
to surround the intense radio emission of the brightest cluster galaxy
(BCG), which often shows non-thermal radio jets and lobes ejected by
the central active galactic nucleus (AGN). The radio-lobe plasma
strongly interacts with the ICM by inflating large X-ray cavities and
triggering the so-called `radio-mode AGN feedback’
\citep[e.g.,][]{Gitti_2012}. The central radio-loud AGN is also likely
to play a role in the initial injection of the relativistic particles
emitting in the mini-halo region. Nonetheless, the diffuse radio
emission of mini-halos is truly generated from the ICM on larger
scales, where the thermal and non-thermal components are mixed, and
can be explained in the framework of leptonic models which envision in
situ particle re-acceleration by turbulence in the cool-core region
\citep[e.g.,][]{Gitti_2002}. Turbulence in cool cores may be generated
by several mechanisms, including the interplay between the outflowing
relativistic plasma in AGN jets/lobes, and sloshing gas motions.

This work is a multi-frequency study of the inner part of Abell~2626
(hereafter A2626), which was included in the first sample of mini-halo
clusters \citep{Gitti_2004}. Its radio morphology, more complex than
that of the standard radio bubbles typically observed to fill X-ray
cavities, represents a challenge to models for the ICM / radio source
interaction in cool cores.
A2626 is a low-redshift \citep[z=0.0553][]{Struble-Rood_1999},
regular, poor cluster \citep[][]{Mohr_1996}, located at RA 23h36m30s,
DEC+21d08m33s and it is part of the Perseus-Pegasus
super-cluster. A2626 has an estimated mass of 1.3$\times10^{15}$
M$_{\odot}$ and a virial radius of 1.6 Mpc \citep[][]{Mohr_1996}.  It
is a cool-core cluster with estimated X-ray luminosity of $1.9 \times
10^{44}$ erg s$^{-1}$ and mass accretion rate of 5 M$_{\odot}$
y$r^{-1}$ \citep[][]{Bravi_2016}.  Its core-dominant (cD) galaxy
IC5338 hosts a pair of optical nuclei with a projected separation of 3.3
kpc, of which only the southern one has a counterpart also in the
radio \citep[][hereafter G13]{Gitti_2013b} and hard X-rays
\citep[][]{Wong_2008} bands \citep[see][Fig. 3]{Wong_2008}.  The
extended radio emission of the cluster has a peculiar morphology
resembling a giant kite, with striking arc-like, symmetric features
whose origin is puzzling \citep[][]{Gitti_2004, Gitti_2013b,
  Kale_2017}. The arcs are collimated structures having largest extent
of $\simeq$70 kpc, steep radio spectrum
\citep[$\alpha<$-2.5\footnote{In this work the radio spectral index
  $\alpha$ is defined such as $S \propto
  \nu^{\alpha}$},][]{Kale_2017}, and no evidence of polarized emission
(G13).  Gitti et al. (2004) argued that the elongated features are
distinct from and embedded in the diffuse, extended radio emission,
which they classified as a radio mini-halo and successfully modeled as
radio emission from relativistic electrons reaccelerated by MHD
turbulence in the cool-core region. On the other hand, the origin and
nature of the radio arcs is still unclear. Their morphology may
suggest that the arcs are radio bubbles, or cluster relics.  However,
Chandra and XMM-Newton observations failed to detect any X-ray cavity
or shock front associated to them, in general showing no clear spatial
correlation between the non-thermal emission of the arcs and the
thermal emission of the cluster \citep[][]{Wong_2008}.

\citet[][]{Wong_2008} argued that the peculiar radio morphology of the
northern and southern arcs may be produced by jet precession triggered
by the reciprocal gravitational interactions of the two cores of the
cD galaxy.
According to this scenario, the relativistic plasma of the arcs was
accelerated in a pair of precessing hot-spots powered by the AGN
located inside the southern core of IC5338.
In particular, if two jets ejected towards the north and south
direction are precessing about an axis which is nearly perpendicular
to the line-of-sight and are stopped at approximately equal radii from
the AGN (at a `working surface'), radio emission may be produced by
particle acceleration, thus originating the elongated structures. The
impressive arc-like, mirror symmetric morphology of these features
highlighted by the high resolution radio images (G13) may support this
interpretation.
On the other hand, the recent discovery of the third and fourth radio
arc to the east-west direction \citep[][]{Gitti_2013b, Kale_2017}
further complicates the picture. They could represent other radio
bubbles ejected in a different direction, similarly to what observed in
RBS~797 \citep{Doria_2012,Gitti_2013a}, but again the absence of any
correlation with the X-ray image disfavors this interpretation. In the
model proposed by \citet{Wong_2008}, they could represent the result
of particle acceleration produced at a working surface by a second
pair of jets ejected to the east-west direction. This interpretation
would imply the existence of radio jets emanating also from the
northeast nucleus of the cD galaxy IC 5338, which however does not
show a radio core.

In order to constrain the jet-precession scenario, we requested new
Karl Jansky Very Large Array (JVLA) observations at 3 GHz and 5.5 GHz to
estimate the spectral index distribution along the arcs and thus infer
detailed information on the radiative age of the plasma in the radio
arcs.
According to the jet-precession model, due to the gradual formation of
the structures, the plasma along the arcs should exhibit a monotonic
trend from one end to the other in the radiative age distribution, and
thus also in the spectral index. The differences in radiative age
between the opposite ends of each arc would represent a measure of the
time required to create the arcs, and therefore it would also provide
an estimate of the precession time of the jets.

We adopt a $\mathrm{\Lambda CDM}$ cosmology with $\mathrm{H_{0}=70}$
km $\mathrm{s^{-1}Mpc^{-1}}$, $\Omega_{M} = 1 - \Omega_{\Lambda} =
0.3$. The cluster luminosity distance is 232 Mpc, leading to a
conversion of 1 arcsec = 1.1 kpc\footnote{{\ttfamily
    https://ned.ipac.caltech.edu/}}.

\section{Observation and Data Reduction}

We performed new observations of the radio source A2626 at 3 GHz
and 5.5 GHz with the JVLA in C-configuration (see Table \ref{obs.tab} for details
regarding these observations). In all observations the source 3C~48
(J0137$+$3309) was used as the primary flux density calibrator, while
the sources J0016$-$0015 and 3C 138 (J0521$+$1638) were used as secondary
phase and polarization calibrators, respectively. Data reduction was
done using the NRAO Common Astronomy Software Applications package
(CASA), version 4.6.  As a first step we carried out a careful editing
of the visibilities. In particular, the 3 GHz dataset required an
accurate flagging to remove every radio frequency interference (RFI),
that are quite common in this radio band. Overall, we removed about
10$\%$ of the visibilities in the 5.5 GHz dataset and 20$\%$ in the 3
GHz one, by repeating manual and automatic flagging with the modes
{\ttfamily MANUAL} and {\ttfamily RFLAG} of the CASA task {\ttfamily
  FLAGDATA}. Due to the visibility loss, the bandwidth of the 3 GHz
observation decreased from 2.0 GHz (2.0-4.0 GHz) to 1.6 GHz (2.4-4.0
GHz), thus moving the central frequency from 3.0 GHz to 3.2 GHz. We
performed a standard calibration procedure\footnote{e.g., see the
  Tutorials,\\ {\ttfamily https://casaguides.nrao.edu/index.php/}} for
each dataset, and the target was further self-calibrated.

Due to the presence of several bright radio sources in the field of
view, we carried out the imaging procedure, with the task {\ttfamily
  CLEAN}, on a $32'\times32'$ region centered on the cluster, in order
to remove as best as possible their secondary lobes. To this purpose,
we used {\ttfamily gridmode=WIDEFIELD} to parametrize the curvature of
the sky regions far from the phase center. We also used the MS-MFS
algorithm \citep[][]{Rau_2011} by setting a two-terms approximation of
the spectral model ({\ttfamily nterms=2}) and the multi-scale clean
({\ttfamily multiscale=[0,5]} on the point-like and beam scales) to
reconstruct as best as possible the faint extended emission.

\begin{table}
\begin{center}
  \caption{ New JVLA data analysed in this work (project code:14B-022)  }
\begin{tabular}{lcc}
\hline
\hline
  PI: Dr. Myriam Gitti & 5.5 GHz  & 3 GHz       \\
~& [C-band] & [S-band]   \\
\hline
~&~&~\\
Observation Date & 19-Oct-2014 & 14-Dec-2014  \\
Frequency Coverage (GHz) & 4.5-6.5 & 2.0-4.0      \\
Array Configuration & C & C \\
On source Time  & 44 m & 63 m \\
\hline
\hline
\label{obs.tab}
\end{tabular}
\vspace{-0.15in} \tablefoot{ A JVLA standard dataset is parted in 16
  spectral windows (spw) with a bandwidth of 128 MHz. Each spw is, in
  turn, divided in 64 channels with a channelwidth of 2 MHz.  }
\end{center}
\vspace{-0.2in}
\end{table}

\section{Results}

We report here the most relevant results of our analysis. For each
observing band, we produced three different types of maps by varying
the relative weight between short and long baselines.  The {\ttfamily
  UNIFORM} maps, obtained by setting {\ttfamily weighting=UNIFORM,
  nterms=1}, give a uniform weight to all spatial frequencies, thus
enhancing angular resolution. The {\ttfamily NATURAL} maps, obtained
by setting {\ttfamily weighting=NATURAL, nterms=1}, give more weight
to low spatial frequencies thus degrading angular resolution but at
the same time maximizing the sensitivity to the diffuse, extended
emission sampled by short baselines. Finally, the {\ttfamily ROBUST 0}
maps, obtained by setting {\ttfamily weighting=BRIGGS, ROBUST=0,
  nterms=2}, have resolution and sensitivity half-way between
{\ttfamily UNIFORM} and {\ttfamily NATURAL}. We report the maps in
Fig. \ref{cmaps.fig} (5.5 GHz) and Fig. \ref{smaps.fig} (3 GHz),
whereas the flux density values measured on each map are reported in
Tab. \ref{flulli.tab}.  Typical amplitude calibration errors are at
3$\%$, therefore we assume this uncertainty on the flux density
measurements. In the following analysis we assume the 7 $\times$ rms
levels at 1.4 GHz (shown in red in Fig. 2, middle panel) as reference
contours for the size and position of the arcs and of the unresolved
core, as they best trace the morphology of the features seen at 1.4
GHz (G13).

\subsection{5.5 GHz maps}

The published map at 4.8 GHz (G13) does not show diffuse radio
emission neither discrete features like the radio arcs seen at 1.4
GHz. The new observations presented in this work were performed with a
more compact configuration of the array and with the larger receivers
band of the JVLA, thus reaching an unprecedented high
sensitivity. This allowed us to detect faint extended emission around
IC5338 and IC5337 for the first time at 5.5 GHz.

\begin{figure}
\hspace{10mm}
\centering
 \begin{minipage}[l]{.5\textwidth}
  \centering
   \includegraphics[width=1\textwidth]{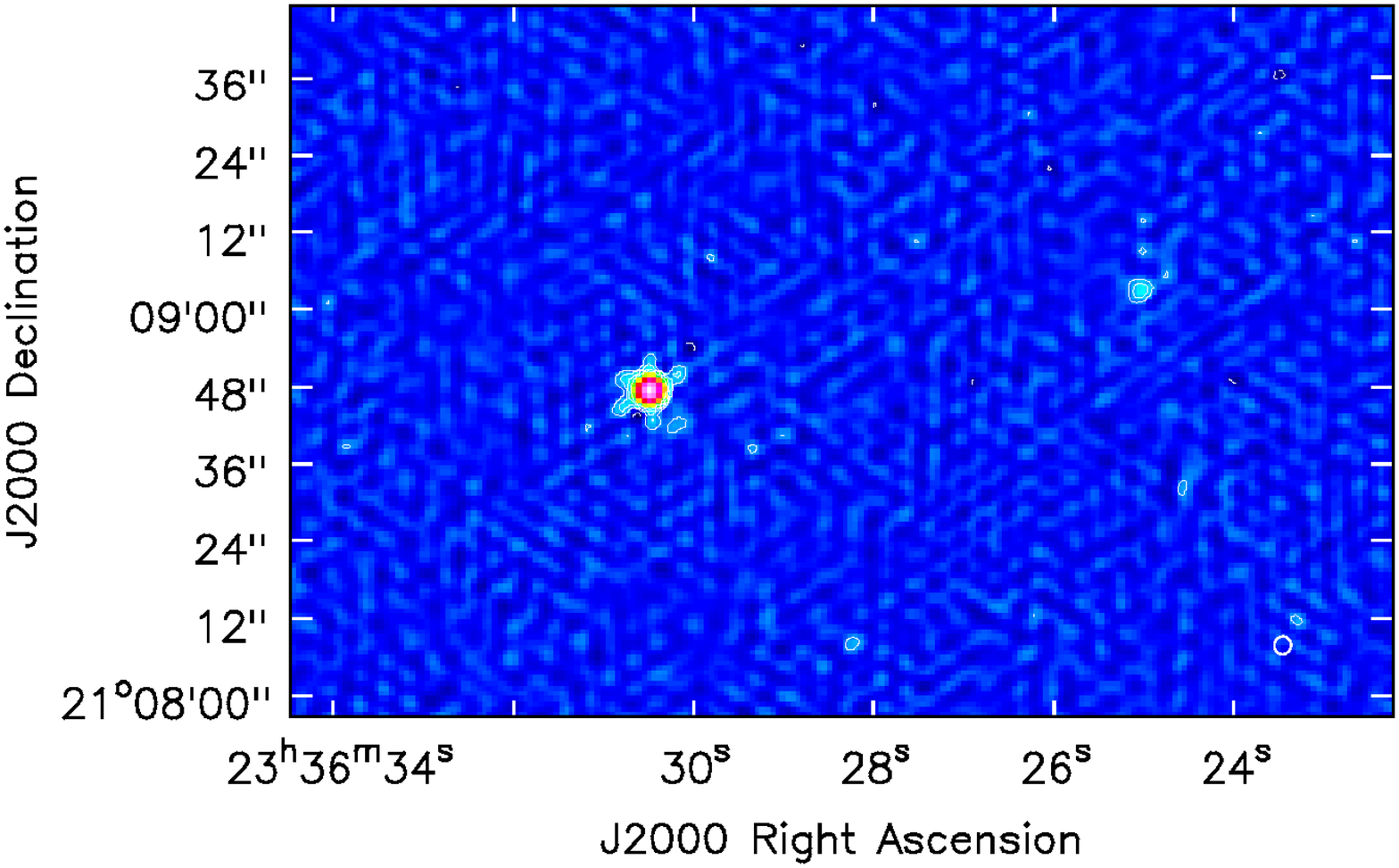}
   \end{minipage}%
 \centering
 \hspace{40mm}
 \begin{minipage}[r]{.5\textwidth}
  \centering
   \includegraphics[width=1\textwidth]{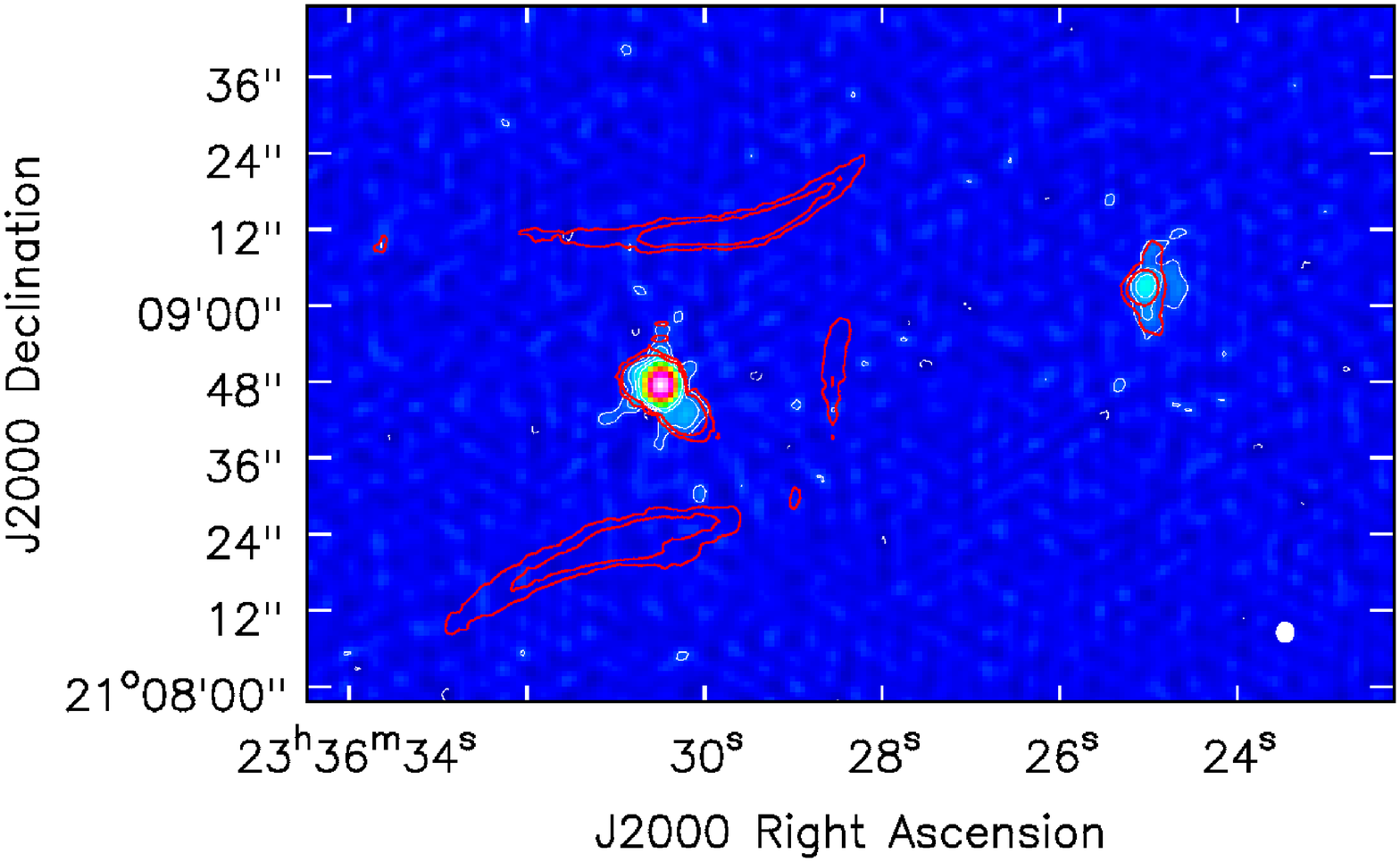}
   \end{minipage}%
  \hspace{-40mm}%
   \begin{minipage}[r]{.5\textwidth}
    \centering
\includegraphics[width=1\textwidth]{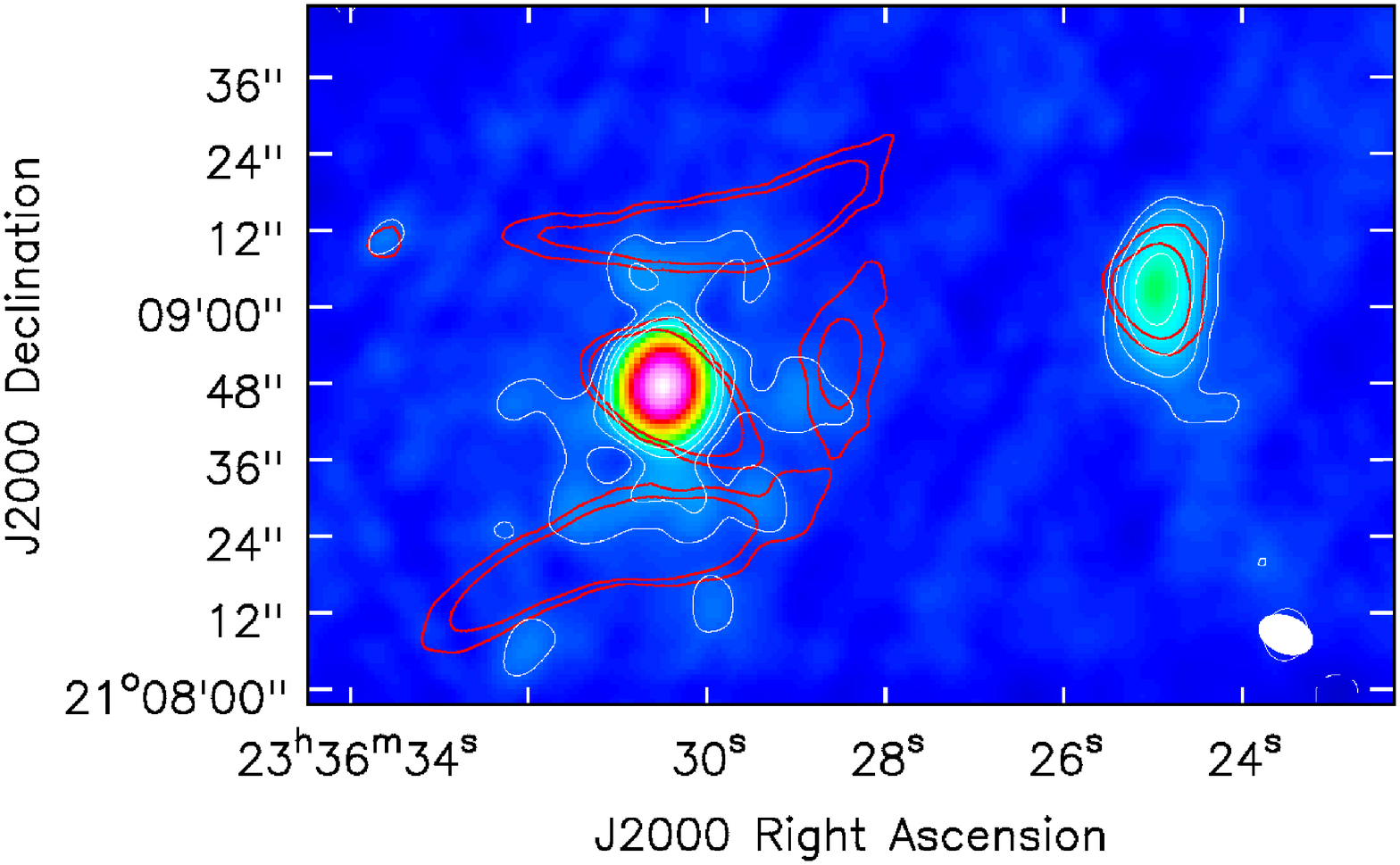}
    \end{minipage}
    \caption{\label{cmaps.fig} {\it Top panel : } 5.5 GHz map
      ({\ttfamily UNIFORM}) at a resolution of $2''.8 \times 2''.7$
      with rms noise of 16.1 $\mu$Jy beam$^{-1}$.  {\it Middle panel :
      } 5.5 GHz map ({\ttfamily ROBUST 0}) at a resolution of $3''.2
      \times 2''.8$ with rms noise = 8.0 $\mu$Jy beam$^{-1}$.  {\it
        Bottom panel : } 5.5 GHz map ({\ttfamily NATURAL,
        UVTAPER=[0,20]}) at a resolution of $4''.9 \times 4''.4$ with
      rms noise = 8.6 $\mu$Jy beam$^{-1}$. In all panels the white
      contours levels are -3, 3, 6, 12, 24 $\times$ rms noise, the red
      ones are the 7, 14 $\times$ rms of 1.4 GHz maps produced with
      comparable resolution from G13 data. The rms levels of the red
      contours are 15.4, 21.9 $\mu$Jy beam$^{-1}$ in middle and bottom
      panel, respectively.}
    \end{figure}

We report the 5.5 GHz maps in Figure \ref{cmaps.fig}. The {\ttfamily
  UNIFORM} (top panel) and the {\ttfamily ROBUST 0} (middle panel)
maps, which have a resolution of $2''.8 \times 2''.7$ and $3''.2
\times 2''.8$, do not show extended emission around IC5338, except for
the jet-like feature at south-west that was yet observed at 1.4 GHz by
G13. On the other hand, in the {\ttfamily ROBUST 0} map the close
source on the left, IC5337, exhibits an extended emission in addition
to the radio core already detected by G13. Due to its optical
properties indicating that the galaxy is leaving a trail of cold gas,
IC5337 has also been classified by as a jellyfish galaxy
\citep[][]{Poggianti_2016}. Therefore, the disrupted radio morphology
agrees with previous dynamical studies of the cluster
\citep[][]{Mohr_1996} indicating that IC5337 is moving toward IC5338.

In the bottom panel of Fig. \ref{cmaps.fig} we show the {\ttfamily
  NATURAL} map at a resolution of $4''.9 \times 4''.4$, obtained by
further imposing a UV tapering to 20 k$\lambda$ to maximize the
sensitivity to the short spatial frequencies. In this
high-sensitivity, low-resolution image we detected extended emission
around IC5338. The emission region extends up to 27$''$ ($\sim$ 30
kpc) and is located inside the radio arcs known at 1.4 GHz (G13). We
measure a total flux densities of $S_{\rm 5.5,NAT} =$ 8.2 $\pm$ 0.2,
of which 6.2$\pm$0.2 mJy are contributed by the core, thus leaving an
estimate of 2.0$\pm$0.3 mJy related to the extended emission. The
nature of this radio emission is uncertain. On the one hand, due to
its position, it may be contributed by the emission of the radio arcs
and of the inner AGN jets. On the other hand, it may indicate the
presence of a diffuse radio mini-halo.

We note that the core flux density measured in these new 5.5 GHz maps
differs from that reported by G13 in the same band, which is
9.6$\pm$0.3 mJy. We accurately checked the flux calibration procedure
in both observations finding no trivial problems, therefore this
$\approx 40\%$ flux difference may indicate variability, and therefore
activity, of the IC5338 core. By observing hard X-rays emission from
its core, \citet{Wong_2008} suggested that IC5338 hosts an AGN. The
jet-like features observed at 1.4 GHz by G13 and the likely
variability of the radio emission that we observe are consistent with
this hypothesis.

\subsection{3 GHz maps}

\begin{figure}
\hspace{10mm}
\centering
 \begin{minipage}[l]{.5\textwidth}
  \centering
   \includegraphics[width=1\textwidth]{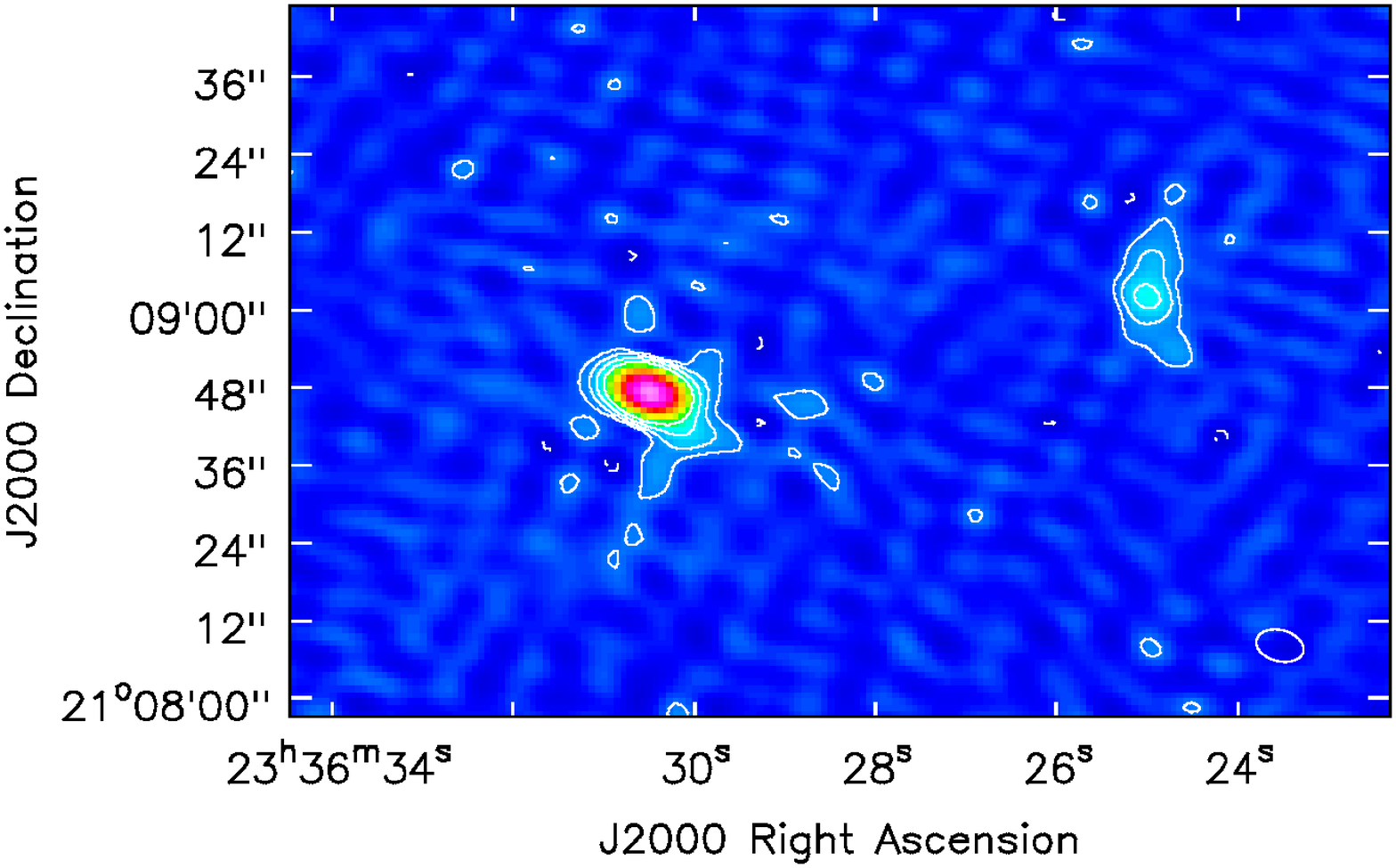}
   \end{minipage}%
 \centering
 \hspace{40mm}
 \begin{minipage}[r]{.5\textwidth}
  \centering
   \includegraphics[width=1\textwidth]{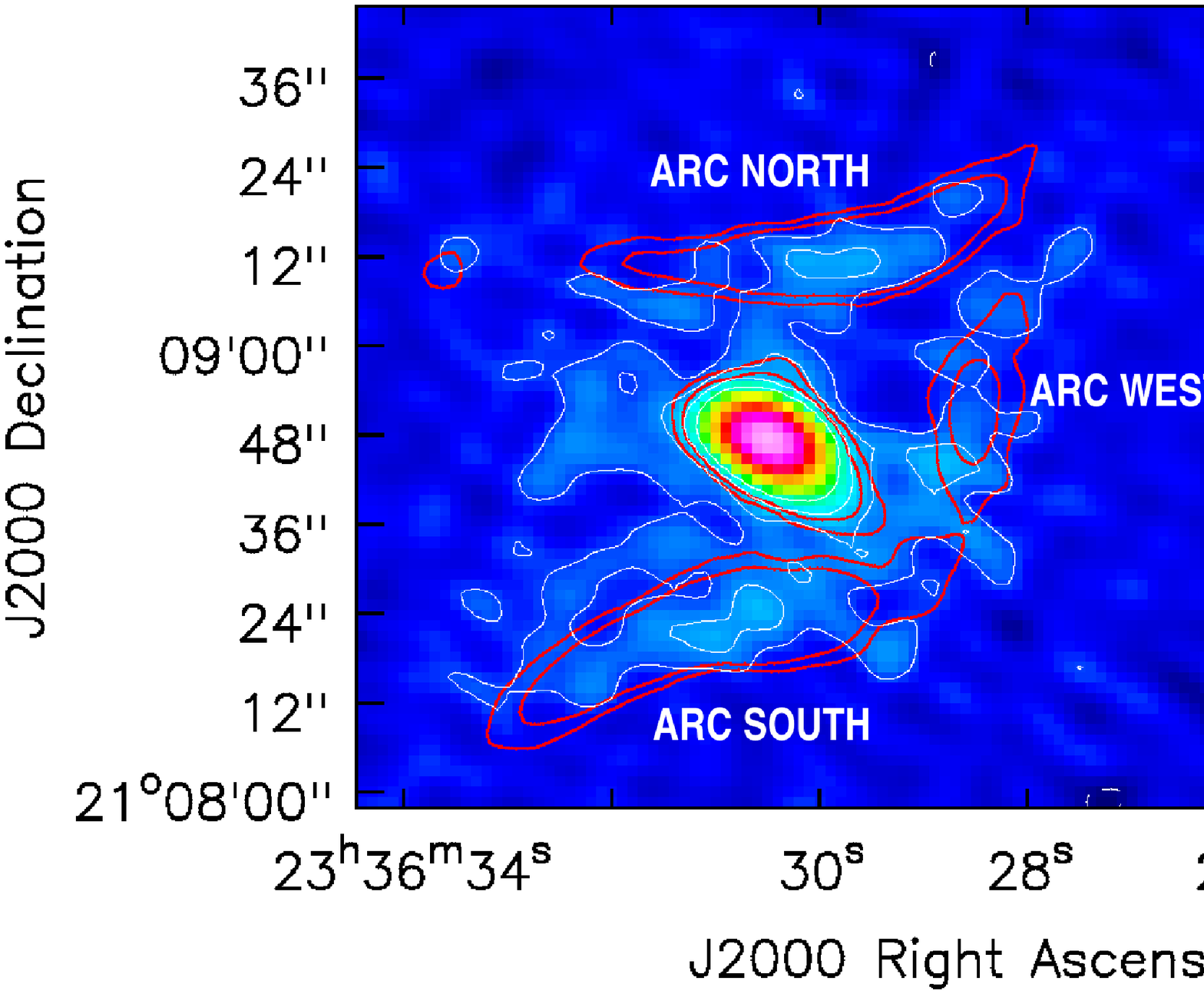}
   \end{minipage}%
  \hspace{-40mm}%
   \begin{minipage}[r]{.5\textwidth}
    \centering
\includegraphics[width=1\textwidth]{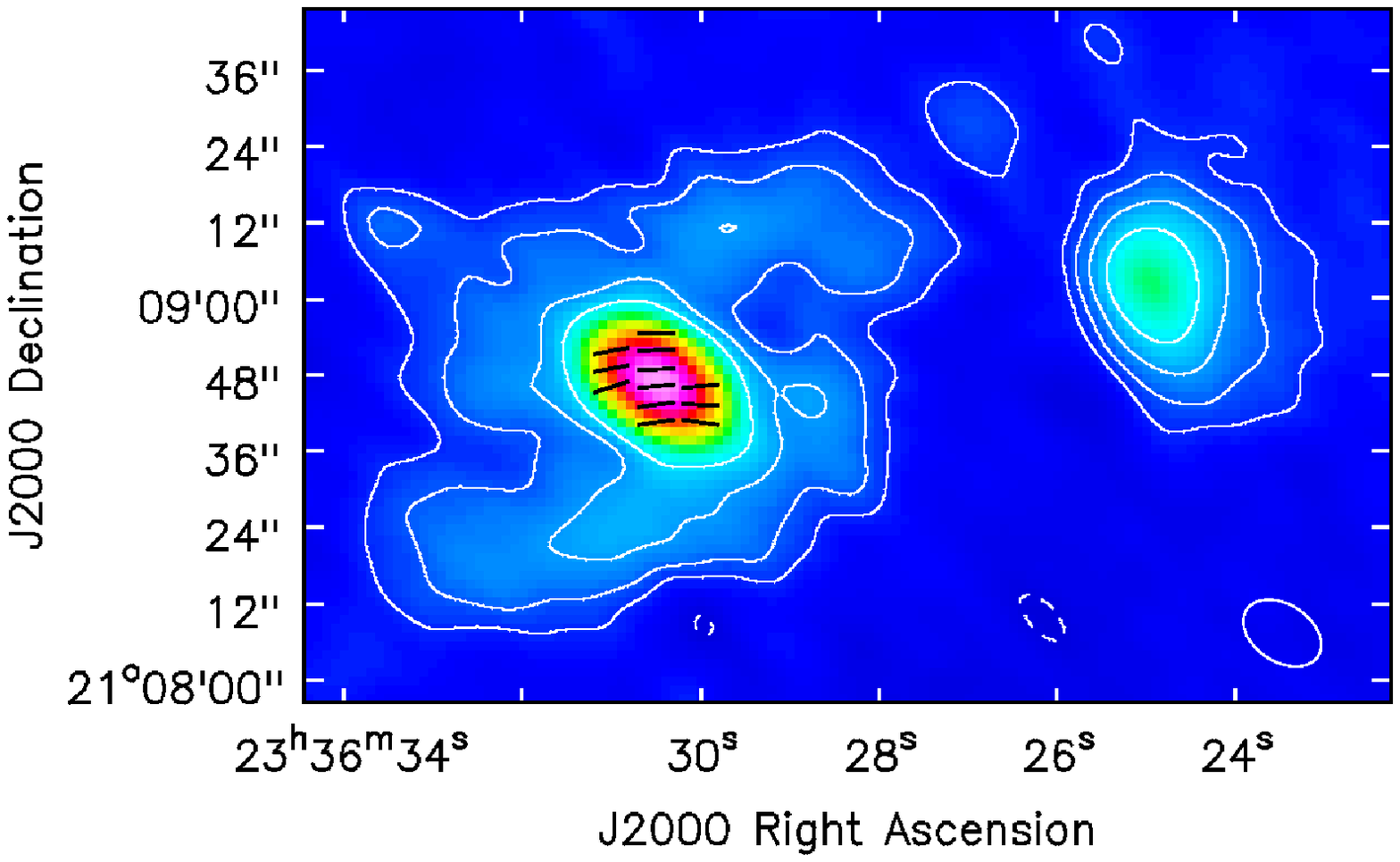}
    \end{minipage}
    \caption{\label{smaps.fig} {\it Top panel : } 3 GHz map
      ({\ttfamily UNIFORM}) at a resolution of $7''.5 \times 8''.5$
      with rms noise = 40.1 $\mu$Jy beam$^{-1}$.  {\it Middle panel :
      } 3 GHz map ({\ttfamily ROBUST 0}) at a resolution of $8''.7
      \times 5''.8$, and rms noise = 20.4 $\mu$Jy beam$^{-1}$. {\it
        Bottom panel : } 3 GHz map ({\ttfamily NATURAL,
        UVTAPER=[0,20]}) at a resolution of $13''.1 \times 8''.5$, and
      rms noise = 18.5 $\mu$Jy beam$^{-1}$. In black there are the
      polarization vectors. In all panels the white contours levels
      are -3, 3, 6, 12, 24 $\times$ rms noise in all maps, the red ones
      are the 7, 14 $\times$ rms of 1.4 GHz maps made with comparable
      resolution.The rms levels of the red contours are 21.9 $\mu$Jy
      beam$^{-1}$.}
\end{figure}

The 2.0-4.0 GHz band is the new radio window observed with the
JVLA. This is also the transmission band of the communication
satellites, so it is populated by radio interferences which may
jeopardize the observations. The maps we present here have been
produced after an accurate editing of the target visibilities. The
{\ttfamily UNIFORM} map (Fig. \ref{smaps.fig}, top panel) exhibits
only the emission of IC5338 and the extended emission of IC5337. The
{\ttfamily ROBUST 0} map at a resolution of $8''.7 \times 5''.8$
(Figure \ref{smaps.fig}, middle panel) shows several patches of
extended emission. By superposing the 1.4 GHz contours by G13 (in
red), we confirm that the new features detected in this band are the
radio arcs already seen at 1.4 GHz.  Remarkably, we identify a feature
to the east which resembles the diffuse emission detected by G13 (see
their Fig. 3), but it does not coincide entirely with the eastern arc
discovered by \citet[][]{Kale_2017}.
Due to the low resolution of this map, the inner part of IC5338 is unresolved.

For this band we produced also a polarization intensity map. We
combined the stokes components, Q and U, of the emission to compute
the vector and total intensity maps of linear polarization.  We set
the weights to {\ttfamily NATURAL} to improve the signal-to-noise
ratio (SNR) of the arcs.  The {\ttfamily NATURAL} map in
Fig. \ref{smaps.fig}, bottom panel, shows an overlay of polarization
vector and total intensity contours. We observe only a small percentage
($\sim2\%$) of polarized emission from the core, whereas we do not observe
polarized emission from the arcs. By comparing the rms of the
polarized emission map with the peak flux density of the arcs, we
estimate an upper limit for the polarized emission fraction of the
arcs of $\sim30\%$\footnote{We note that the low resolution map is
  affected by the beam depolarization effect, so high-resolution
  observations are needed to confirm the lack of polarized emission
  from the radio arcs at 3 GHz.}.

\begin{table}
\begin{center}
  \caption{ Flux density values of the radio sources of A2626,
    measured on the maps of Fig. \ref{cmaps.fig}-\ref{smaps.fig}
    inside the 3 $\times$ rms contours of each map.}
\begin{tabular}{ccc}
\hline
\hline
Weighting&Component&Flux density\\
&&[mJy]\\
\hline
&\textbf{5.5 GHz maps} (Fig. \ref{cmaps.fig})&\\
\hline
&Total & 6.7$\pm$0.2\\
 {\ttfamily UNIFORM} &  Core & 6.7$\pm$0.2\\
  & IC 5337 & 0.2$\pm$0.1 \\
&&\\ 
&Total & 7.4$\pm$0.2\\
 {\ttfamily ROBUST 0} &  Core & 6.0$\pm$0.2\\
  & IC 5337 & 0.5$\pm$0.1 \\
&&\\
&Total & 8.2$\pm$0.2\\
 {\ttfamily NATURAL} &  Core+jets & 6.2$\pm$0.2\\
  & IC 5337 & 0.5$\pm$0.1 \\
   \hline
   &\textbf{3 GHz maps} (Fig. \ref{smaps.fig})&\\
   \hline
&Total & 12.1$\pm$0.5\\
 {\ttfamily UNIFORM} &  Core+jets & 11.7$\pm$0.4\\
  & IC 5337 & 1.6$\pm$0.3 \\
&&\\ 
\multirow{6}{*}{{\ttfamily ROBUST 0} }&Total & 14.6$\pm$0.4\\
   &Core+jets & 12.0$\pm$0.4\\
   &Arc Noth&0.4$\pm$0.1\\
   &Arc South&0.5$\pm$0.1\\
   &Arc West&0.5$\pm$0.1\\
   &IC 5337 & 1.5$\pm$0.1 \\
&&\\  
   \multirow{6}{*}{{\ttfamily NATURAL} }& Total & 16.5$\pm$0.5\\
   &Core+jets & 10.9$\pm$0.3\\
   &Arc Noth&0.7$\pm$0.1\\
   &Arc South&0.8$\pm$0.1\\
   &Arc West&0.4$\pm$0.1\\
   &IC 5337 & 2.2$\pm$0.1 \\
   \hline
   \hline
\label{flulli.tab}
\end{tabular}
\vspace{-0.15in} 
\end{center}
\vspace{-0.2in}
\end{table}

\subsection{Spectral Index maps}

By combining the new JVLA observation at 3 GHz with the VLA ones at
1.4 GHz obtained with the VLA in A+B configuration (G13), we produced
a spectral index map of the extended emission of A2626. We produced
the input maps by setting {\ttfamily weighting=UNIFORM, UVRANGE=0-40,
  UVTAPER=0-20, RESTORINGBEAM=$13'' \times 10''$, nterms=1,
  multiscale=[0,5]}.  We improved the sensitivity to the faint,
extended emission of the {\ttfamily UNIFORM} maps by applying a
{\ttfamily UVTAPER} and by enlarging the beam to improve the SNR. The
rms of the 1.4 GHz and 3 GHz maps are 31.9 and 29.1 $\mu$Jy
beam$^{-1}$.

\begin{figure*}
\begin{minipage}{.5\textwidth}
\includegraphics[width=1\textwidth]{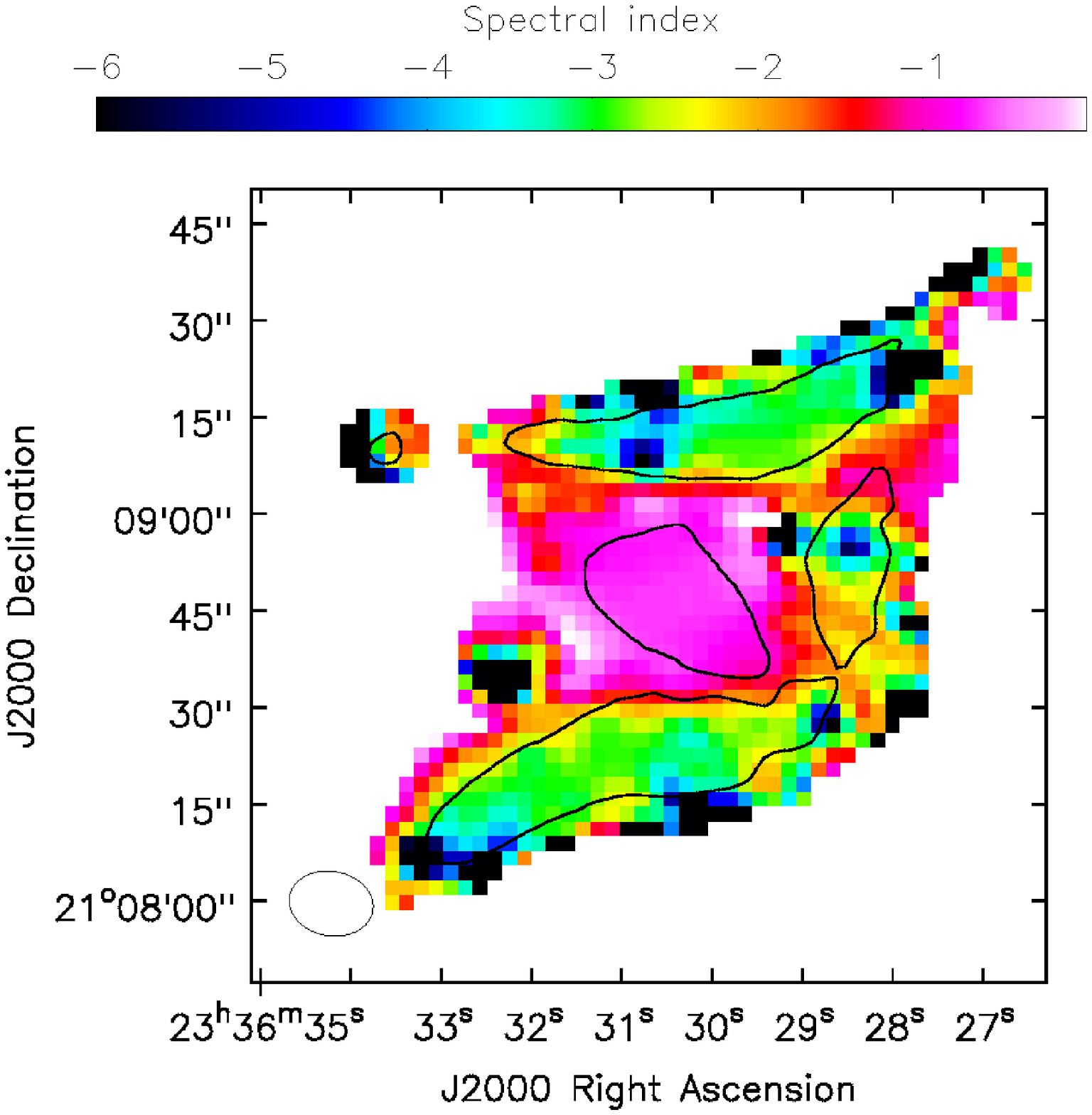}
\end{minipage}
\begin{minipage}{.5\textwidth}
\includegraphics[width=1\textwidth]{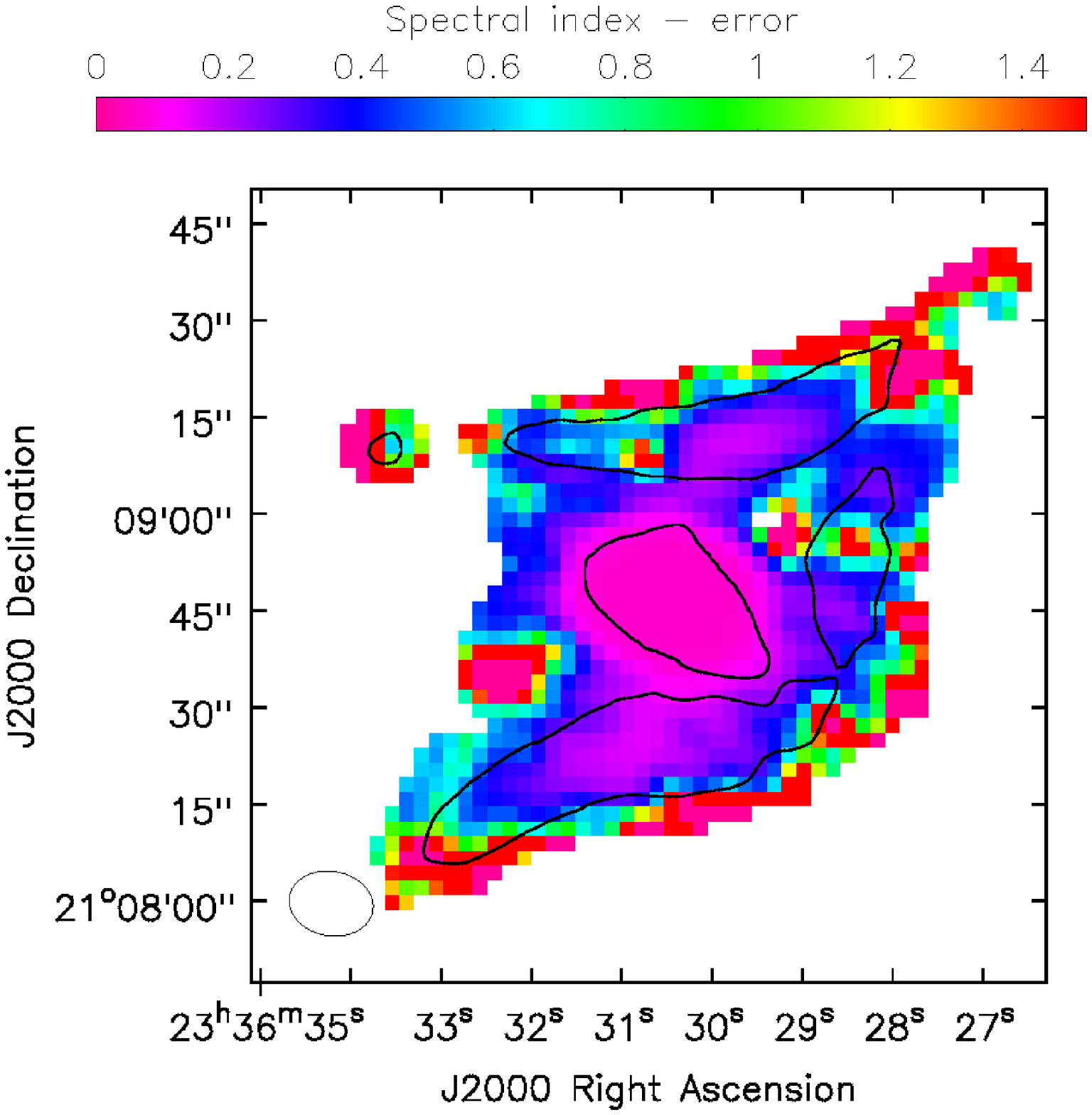}
\end{minipage}
\caption{\label{spix.fig} Bandwidth-corrected spectral index map
  (left) and relative error map (right) obtained from the 1.4 GHz and
  3 GHz dataset. The black contours are the same 1.4 GHz contours
  reported in Fig. \ref{smaps.fig}, middle panel. }
\end{figure*}

In calculating the spectral index value in each pixel of the map, we
excluded regions where the 1.4 GHz brightness is $<3 \times$ rms
level. We further corrected the spectral index map for the bandwidth
effect described by \citet[][]{Condon_2015}. Due to the large receiver
bandwidth of the JVLA, the flux density the we measure on the 3 GHz
maps may not coincide with the flux at 3.2 GHz, which is the
arithmetical center of our band, but instead it may be the flux
density at an equivalent frequency, $\nu_{u}$, defined as:
\begin{equation}
 \nu_{u}=\left[\frac{1}{\alpha+1}\left( \frac{\nu_{max}^{\alpha+1}-\nu_{min}^{\alpha+1}}{\nu_{max}-\nu_{min}}   \right)  \right]^{1/\alpha}
 \label{condon.math}
\end{equation}
where $\nu_{max}$ and $\nu_{min}$ (expressed in GHz) are the ends of
the bandwidth and $\alpha$ is the spectral index.  According to this
relation, $\nu_{u}$ is a function of $\alpha$, which is, in turn, a
function of the frequencies. In order to correct the spectral index
map, which was computed between 1.4 and 3.2 GHz, we derived
numerically the correction for the spectral indices. According to our
bandwidth (2.4-4.0 GHz), we estimated a first set of equivalent
frequencies $\nu_{u}^{i}$ for a set of spectral indices $\alpha_{i}$
between -0.1 and -7.0. We then corrected the spectral indices for the
equivalent frequencies, thus obtaining a new set $\alpha_{i+1}$ that
allowed us to estimate again a new set of equivalent frequencies
$\nu_{u}^{i+1}$. We repeated this cycle until it converges to the
final values $\nu_{u}^{N}$ and $\alpha_{N}$. We then obtained by
numerical fit the relation $\alpha_{N}(\alpha_{i})$ that we finally
applied to the each pixel of the uncorrected spectral index map with
the task {\ttfamily immath} of CASA. With the $\alpha_{N}(\alpha_{i})$
relation we re-scaled also the relative spectral index error map. We
report the corrected spectral index map, and the relative error map,
in Fig. \ref{spix.fig}.

The spectral index map disentangles the extended emission in two major
components, the unresolved core with a flat spectral index\footnote{As
  a word of caution we note that the spectral index in the central
  region may be unreliable due to the core variability, because the
  observations are separated by five years.} and the arcs with steep
spectral index $\alpha<-2.5$, that is consistent with the recent
results by \citet[][]{Kale_2017}. We measured the mean spectral index
inside the 1.4 GHz reference contours (Fig. \ref{smaps.fig}, middle
panel) for the northern, southern and western arcs, finding
$\alpha=-3.2 \pm 0.6$, $-3.0 \pm 0.4$ and $-2.6 \pm 0.6$,
respectively. The unresolved core has a mean spectral index
$\alpha=-0.7\pm0.1$.

\subsection{Radiative age maps}

The continuous radiative loss modifies the spectrum of the emitting
particles by steepening it. So, it is possible to estimate the
radiative age, $t_{rad}$, of the emitting plasma from the break
frequency, $\nu_{br}$, of its synchrotron spectrum and the magnetic
field, $B$, that produces the radio emission \citep[e.g., Eq. 2
of][]{Murgia_1999}\footnote{Note that we are only considering the
  effect of radiative losses on the evolution of the energy of
  electrons}. From this relation, it is possible to demonstrate that
the radiative life-span, for a given $\nu_{br}$, is maximized by a
magnetic field $B_{ml}=B_{CMB}/\sqrt{3}$ $\mu$G, where
$B_{CMB}=3.25(1+z)^{2}$ $\mu$G is the equivalent magnetic field of the
inverse Compton emission and $z$ is the redshift of the radio source.
By integrating numerically the synchrotron emission spectrum, we
derived a $t_{rad}(\alpha)$ relation for a population of particles
with a energy distribution function $n(E)\propto E^{-\Gamma}$ and a
magnetic field $B$ that is uniform and constant over the radio
source. We assumed a $B=B_{ml}=2.1$ $\mu$G as magnetic field, and
$\Gamma$=2.4 for the energy distribution function, that we derived
from the hypothesis of the jet-precession model. If the plasma of the
arcs was initially accelerated in an hot-spot, then it had an initial
spectral index $\alpha=-0.7$ \citep[e.g.][]{Meisenheimer_1997}, and so
$\Gamma=1-2\alpha=2.4$. Therefore, the $t_{rad}$ map exhibits the
upper limit of the time required to the spectrum to steepen from $-0.7$
to the values that we observe in Fig. \ref{spix.fig}. Due to our
assumptions, it was not possible to evaluate the $t_{rad}$ in those
regions which emit with a spectral index flatter than $-0.7$.

\begin{figure*}
\begin{minipage}{.5\textwidth}
\includegraphics[width=1\textwidth]{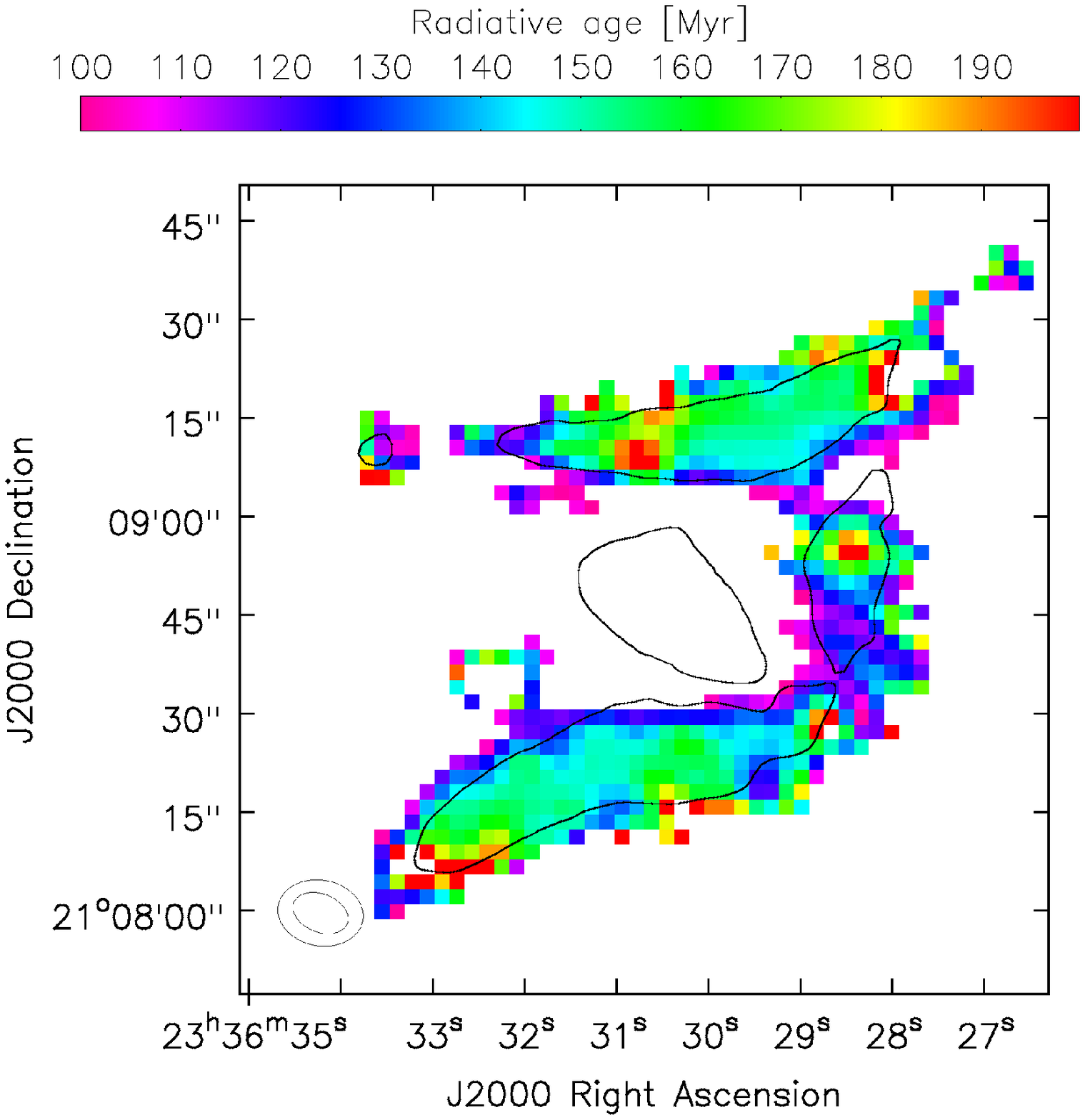}
\end{minipage}
\begin{minipage}{.5\textwidth}
\includegraphics[width=1\textwidth]{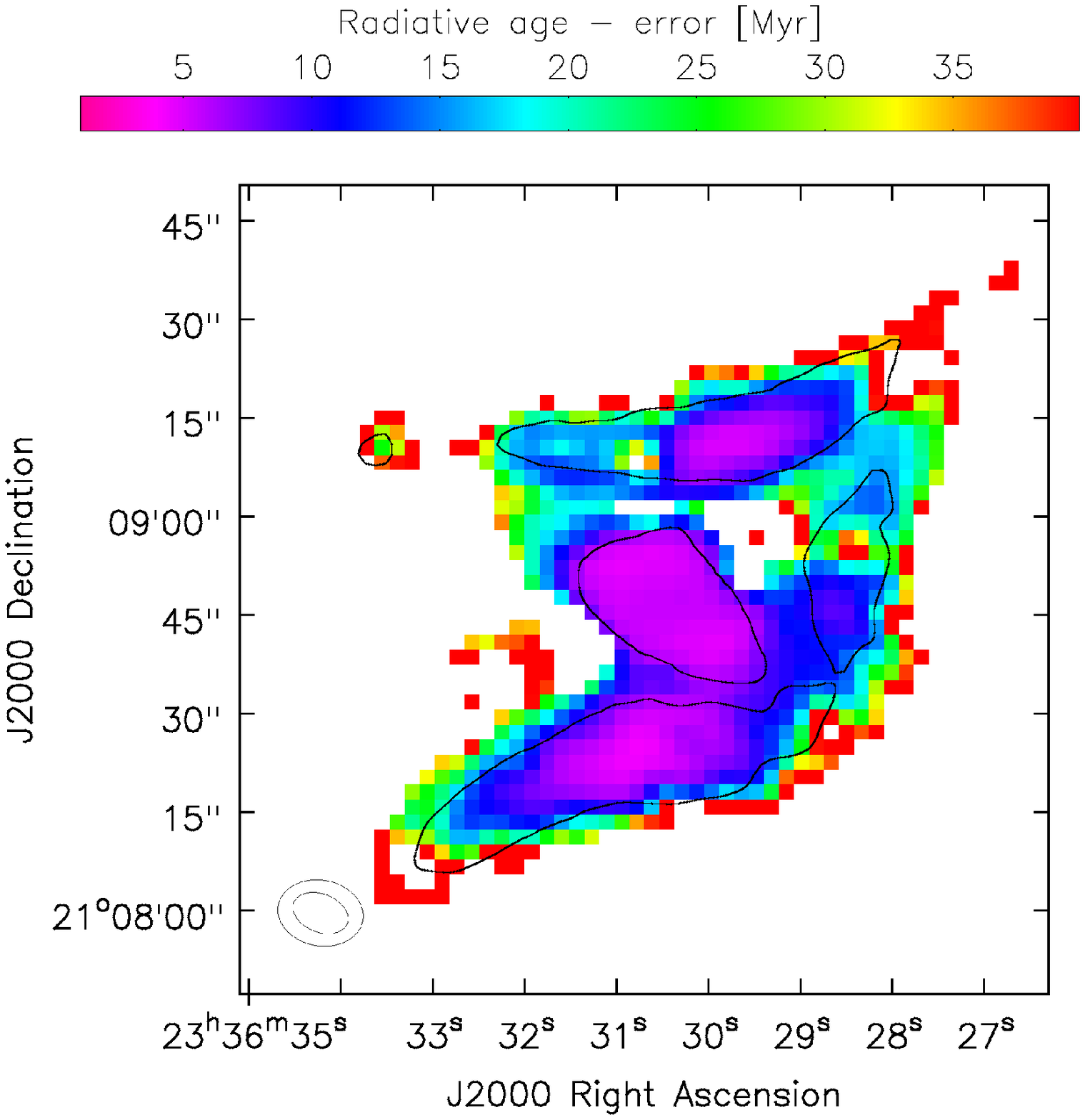}
\end{minipage}
\caption{\label{age.fig} Radiative age map (left) and relative error
  map (right) obtained from Fig. \ref{spix.fig} by assuming $B=2.1$
  $\mu$G and $\Gamma=2.4$. The black contours are the same 1.4 GHz 
  contours reported in Fig. \ref{smaps.fig}, middle
  panel. The map displays the upper limit of the time required to
  steepen the spectral index of the synchrotron emission from $-0.7$ to
  the value reported in Fig. \ref{spix.fig} }
\end{figure*}

We computed the spectral map in Fig. \ref{spix.fig} with the
$t_{rad}(\alpha)$ relation and we obtained the radiative age
map. Moreover, we obtained the relative error map on the radiative age
from the spectral index error map by estimating the radiative age
associated to the upper and lower limits of the spectral index map. We
report the radiative age map and the relative error map in
Fig. \ref{age.fig}.  To obtain an indicative estimate of the mean
upper limit on the radiative age, we averaged the values of the
$t_{rad}$ map inside the 1.4 GHz reference contours
(Fig. \ref{smaps.fig}, middle panel). We measured $t_{rad} =$ 157 Myr,
148 Myr and 136 Myr for the northern, southern and western arcs,
respectively, which can be considered as an estimate of the maximum
time elapsed since the particle acceleration.  The average radiative
age error, $\sigma_{t}$, on those regions is $\sim$12 Myr.

\section{Discussion}

 \begin{figure*}
 \begin{minipage}[l]{.5\textwidth}
 \centering
  \includegraphics[width=1.1\textwidth]{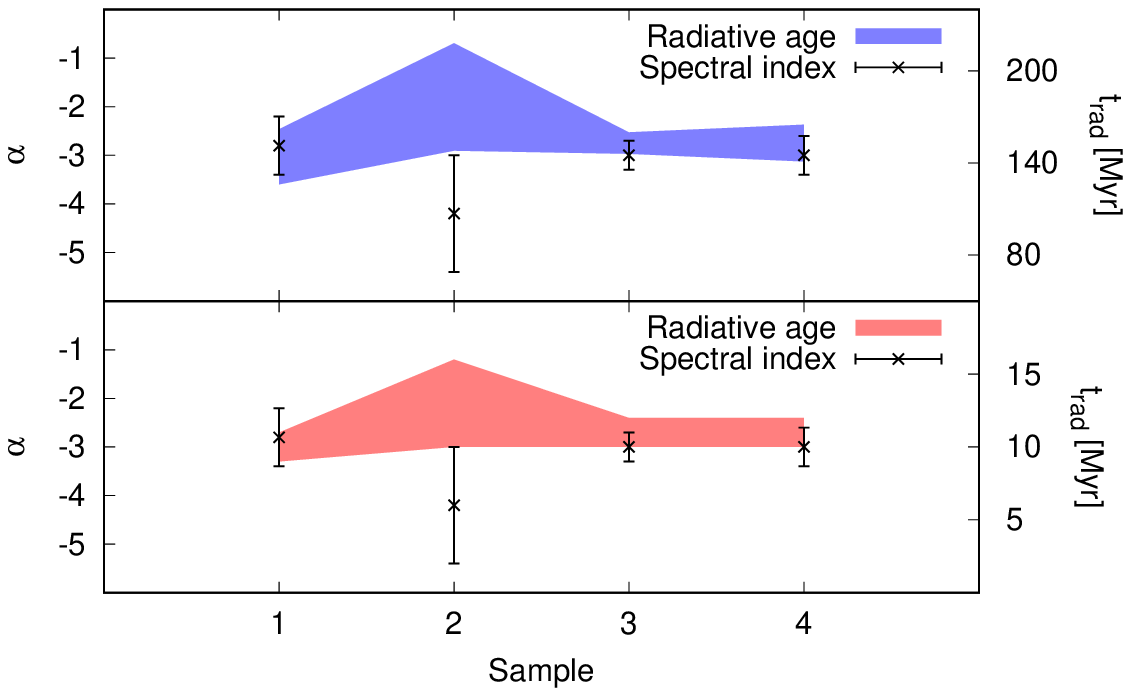}
  \vspace{-1.5cm}
   \end{minipage}%
   \begin{minipage}[r]{.5\textwidth}
    \centering
    \includegraphics[width=1.1\textwidth]{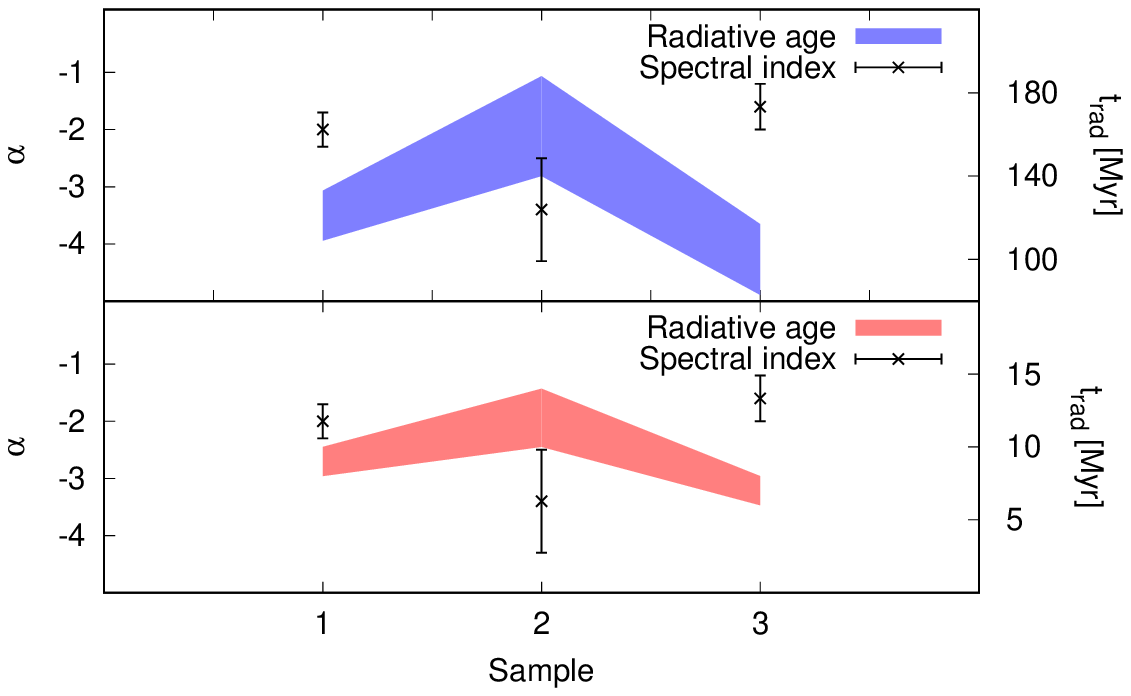}
   \vspace{-1.5cm}
   \end{minipage}
   \begin{minipage}[l]{.5\textwidth}
   \centering
    \includegraphics[width=1.1\textwidth]{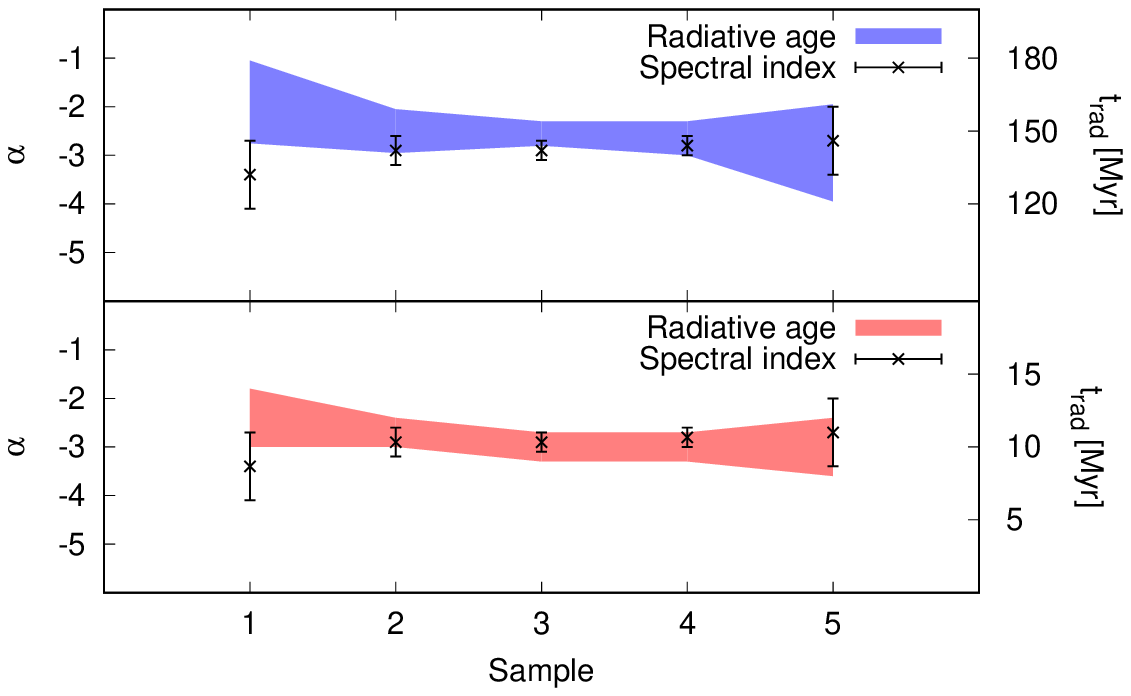}
    \vspace{-1.5cm}
    \end{minipage}
    \begin{minipage}[r]{.4\textwidth}
   \hspace{1cm}
    \includegraphics[width=1\textwidth]{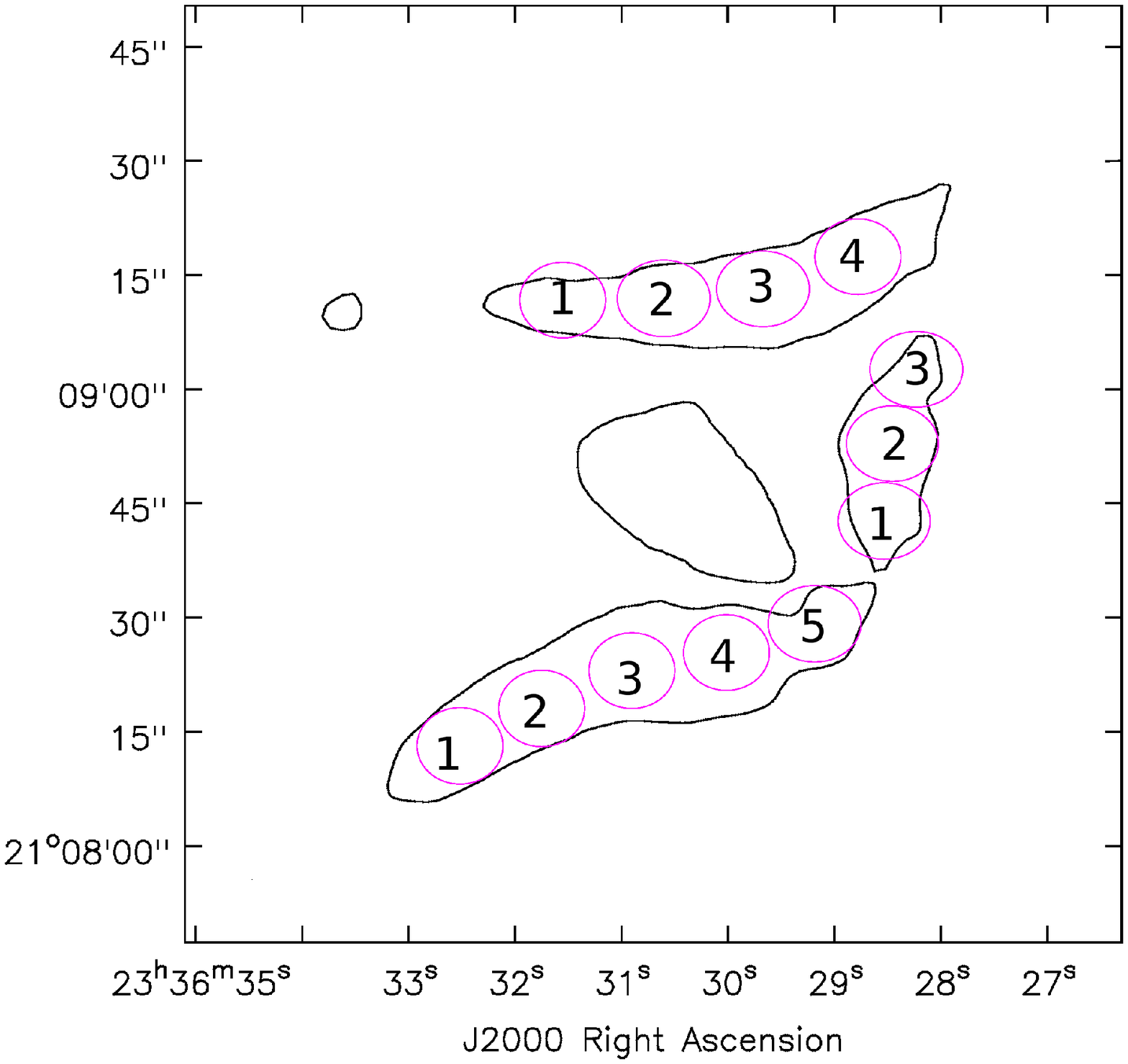}
    \end{minipage}
    \caption{\label{trend.fig} Spectral index and radiative age trend
      along the radio arcs obtained with the sampling method shown in
      the bottom-right panel. On the x-axis is shown the sampling
      numeration (indicated in the bottom right panel), on the left
      y-axis the spectral index scale and on the right y-axis the
      radiative age scale. In each panel the top blue plot displays
      the radiative age estimated by assuming $B=B_{ml}=2.1$ $\mu$G,
      and the bottom red plot by assuming the equipartition magnetic
      field $B=B_{eq}=12$ $\mu$G. {\it Top-left}: Northern arc trends;
      {\it Top-right}: Western arc trends; {\it bottom-left}: Southern
      arc trends; {\it Bottom-right}: Sampling method. The black
      contours are the 1.4 GHz reference levels reported in
      Fig. \ref{smaps.fig}, middle panel;}
\end{figure*}

The new JVLA observations presented in this work add important
information to the general picture of A2626, and allowed us to test
for the first time the jet-precession model.

The spectral index map in Fig. \ref{spix.fig} exhibits that the arcs
are steep-spectrum radio sources, with $\alpha\leq-2.5$. We do not
observe $\alpha\simeq-0.7$ at any of the arcs ends, which should be
the initial spectral index of the plasma, so we argue that the
activity has ceased. The radiative map that we provide
(Fig. \ref{age.fig}) does not show the real age of the arcs, but only
an upper limit of it, because we do not know the real strength of the
magnetic field in the arc regions. However, it is useful to constrain
the time-scale of the jet-precession by considering the difference
between the radiative age at the ends of each arc, $\Delta
t_{rad}$. This can be considered as a measure of the time required by
the jets to cover the arc length, thus it is an estimate of the
time-scale of the precession period of the jets.

According to the jet-precession scenario, the relativistic plasma of
the arcs was accelerated gradually along the structures by a pair of
precessing jets.  Therefore, the spectral index distribution along the
arcs may reflect the gradual formation of the structures, by showing a
monotonic steepening trend from one end to the other, where the
emission with the flattest spectral index comes from the most recently
accelerated plasma. Moreover, the trends along the northern and
southern arcs should be anti-symmetric due to the geometry of the
jet-precession (if the northern arc is created from west to east, then
the southern arc must be created from east to west), and they should
be observable also in the radiative age maps.

 Observationally, the spectral index map (Fig. 3) does not show
  any clear evolution in the spectrum along the arcs, hinting at the
  absence of the age progression expected by the precession model. 
In order to  quantitatively confirm this result, we sampled both
the spectral index and the radiative age maps with the sampling
reported in Fig. \ref{trend.fig} (bottom-right panel) and we measured
the values, with the relative errors, of each sample. The samples are
as large as the beam ($13'' \times 10''$) and cover the area delimited
by the 1.4 GHz reference contours shown in Fig. \ref{smaps.fig},
middle panel. We report the result of our analysis on each arc in
Fig. \ref{trend.fig}.

We estimated the radiative age trend also by assuming the reference
equipartition magnetic field. From the formula provided by
\citet[][ Eq. 22]{Govoni-Feretti_2004}, by assuming that the spectral index
of the arcs at $\leq$100 MHz is $\sim -0.7$\footnote{The classical
  equipartition formula \citep{Pacholczyk_1970} is obtained for a
  power law energy distribution of electrons. However, we detect a
  synchrotron spectrum whose properties are affected by radiative
  ageing.  In order to correct for this effect, we estimated the
  spectrum of the arcs at lower frequencies, where ageing is not
  important and the assumption of power-law energy distribution is
  satisfied. In particular, we calculated the emission at frequency
  100 MHz by re-scaling the spectrum measured at 1.4 GHz according to
  the synchrotron spectrum with a break frequency derived from the
  spectral slope between 1.4 GHz and 3.0 GHz, obtaining an estimated
  flux density at 100 MHz of $\sim$260 mJy.}, that the arcs have a
cylindrical volume and that the energy in electrons and protons is
equally distributed ($k$=1), we estimated a equipartition field
strength of $B_{eq} = 12$ $\mu$G.  Depending on the magnetic field,
the mean radiative age of the arcs decreases from $\sim$140 Myr to
$\sim$10 Myr, whereas the mean error on it decreases from $\sim$10 Myr
to $\sim$1 Myr. Due to the uncertainties in the assumptions made to
derive the equipartition magnetic field, in the following discussion
we consider the estimates of radiative age upper limits provided by
$B_{ml}$.

The plots in Fig. \ref{trend.fig} show that the radio arcs do not
exhibit significant spectral index or radiative age trends. By
considering the uncertainties of our measures, we derived an estimate
of the maximum difference between the radiative time scale
$\Delta t_{rad}$ of 22 Myr, 26 Myr and 22 Myr for the northern,
southern and western arcs, respectively. This corresponds to an
estimate of the upper limit on the time required by the jets to cover
the arc length.

In order to test the jet-precession model, we compared our result with
a theoretical estimate of the precession period. \citet[][]{Wong_2008}
argued that the jet-precession may be triggered by the mutual
gravitational interactions of the optical cores of IC5338. The
precession period, $\tau_{prec}$, of the jets of a binary system in
the center of a galaxy cluster can be estimated with the relation
proposed by \citet[][]{Pizzolato_2005}. We re-scaled the relation that
they provide according to the parameters that we observe:

\begin{equation}
  \tau_{prec}=1.6\cdot10^{4} \left( \frac{M}{10^{8}M_{\odot}} \right)^{-1/2}\left( \frac{a}{1\text{ kpc}} \right)^{3}\left( \frac{a_{D}}{1\text{ pc}} \right)^{-3/2}\frac{(1+q)^{1/2}}{q\text{ }cos\theta} \text{Gyr}
\end{equation}
where $M$ and $q$ are, respectively, the sum and the ratio of the
masses of the black holes, $a$ is their distance, $a_{D}$ is the
radius of the accretion disk and $\theta$ is the angular difference
between the accretion disk and the plane of the orbit of the black
holes.  In the case of A2626, the masses of the objects and the
accretion disk radius are unknown, so we assumed a mass of $10^{8}$
M$_{\odot}$ (so $M=$2$\times 10^{8}$ M$_{\odot}$, $q$=1), that is the
typical mass of the supermassive black holes (SMBH) located at the
center of cD galaxies, and the typical accretion disk radius $a_{D}=1$
pc. The observed projected distance between the cores, which is a
lower limit to their real distance, is $a$=3.3 kpc. Moreover, also
$\theta$ is unknown, therefore we assume for simplicity that the
accretion disk is on the same plane of the orbit ($\theta=0$). From
these assumptions, we estimated a lower limit for the precession time
$\tau_{prec}\gtrsim 10^{5}$ Gyr, that is not in agreement with the
$\Delta t_{rad}$ that we measure. This estimate is also much greater
than the age of the universe, thus challenging the hypothesis that the
two nuclei are currently producing the jet-precession. More
specifically, from the formula proposed by \citet[][]{Pizzolato_2005},
the distance between the two cores that admits a precession period
$\tau_{prec}\simeq \Delta t_{rad}$ is:
 \begin{equation}
   a\leq10 \left( \frac{\tau}{20 \text{Myr}}\right)^{1/3}\left( \frac{M}{10^{8} \text{M}_{\odot}}\right)^{1/6}\left( \frac{a_{d}}{1 \text{pc}}\right)^{1/2}\left( \frac{qcos\theta}{(1+q)^{1/2}}\right)^{1/3}\text{ pc} 
 \end{equation} 
 Consequently, at the epoch of particle acceleration (corresponding to
 $t_{rad} \approx 150$ Myr ago) the two cores had to have a separation
 of just $\approx$ few pc. Alternatively, the precession should have
 been generated by the interactions with a secondary SMBH which is yet
 undetected (likely not active), rather than with the northern optical
 core of IC5338.
 Therefore, we argue that the dynamical process behind the origin of
 the radio arcs may be more complex than the scenario initially
 proposed by \citet[][]{Wong_2008}. At the same time, however, the
 lack of a clear optical core candidate makes the precession model
 less appealing, and other scenarios for the origin of the radio arcs
 should be investigated in the future.

 We finally note that, by combining the results of our spectral
 analysis with the spectral maps provided by \citet[][]{Kale_2017}, we
 may argue that the radio spectrum of the arcs has a steep spectral
 index $<-2.5$ from 610 MHz to 3 GHz, and this is not in agreement
 with the expected evolution of the radio spectrum of the hot-spots.
 The AGN-driven nature of the radio arcs through jet precession may be
 further challenged by the fact that we observe four arcs, which would
 require the presence of {\it two pairs} of radio jets, powered by
 {\it two radio cores} inside IC5338. However, only one radio AGN
 (with jets) is currently observed, whereas the other one remains
 undetected (although it is still possible that it was active in the
 past and has recently shut down).

\section{Summary and Conclusion}

We presented the results of new JVLA observations of A2626 at 3 GHz
and 5.5 GHz, which were requested in order to test the jet-precession
model for the origin of the radio arcs. We achieved the first
detection at high frequency of an extended component of radio emission
at 5.5 GHz and we observed for the first time the radio arcs at 3
GHz. Moreover, we observed the extended emission of IC5337 at 3 GHz
and 5.5 GHz. 

By combining the archival 1.4 GHz observation (G13) with the new one
at 3 GHz, we produced new spectral index maps that allowed us to
disentangle the extended emission of the cluster in two main
components -- the inner, flat-spectrum emission
($\alpha \simeq -0.7$), and the ultra-steep spectrum
($\alpha \leq -2.5$) arcs. In order to find constraints for the
jet-precession model, we converted the spectral index map into
radiative age maps and we studied the trends of spectral index
and radiative time along the arcs.  We found that the radio
  arcs do not exhibit significant spectral index or radiative age
  evolution. By considering the uncertainties on our measures, we
  estimated an upper limit of the precession period
  $\Delta t_{rad}\leq 26$ Myr.  We also estimated a mean radiative age
  of $t_{rad} \leq 150$ Myr, that may be considered as a measure of
  the time elapsed since the hot-spot ceased to accelerate the
  particles. 

We then compared our results with a theoretical precession period
estimated for the current kinematics of the IC5338 system, finding a
disagreement. Therefore, we argue that the dynamics of the cores that
produced the jet-precession and created the radio arcs may be more
complex than the scenario proposed by \citet[][]{Wong_2008}. On the
other hand, our results put strong constraints for every future model
which relates the origin of the radio arcs to the activity of the
inner AGN, or indicate a different origin of the radio arcs.

\section*{Acknowledgments}

We thank the referee for the prompt response and constructive comments.
AI thanks R. Paladino for helpful advices during the data reduction
with CASA.  We thank A. Bonafede, M. Bondi and D. Dallacasa for useful
discussions. MG, GB, GG, LF acknowledge partial support from PRIN-INAF
2014. 

\bibliographystyle{aa}
\bibliography{bibliography.bib}

\begin{thebibliography}{21}
\expandafter\ifx\csname natexlab\endcsname\relax\def\natexlab#1{#1}\fi

\bibitem[{{Bravi} {et~al.}(2016){Bravi}, {Gitti}, \& {Brunetti}}]{Bravi_2016}
{Bravi}, L., {Gitti}, M., \& {Brunetti}, G. 2016, \mnras, 455, L41

\bibitem[{{Brunetti} \& {Jones}(2014)}]{Brunetti-Jones_2014}
{Brunetti}, G. \& {Jones}, T.~W. 2014, International Journal of Modern Physics
  D, 23, 30007

\bibitem[{{Condon}(2015)}]{Condon_2015}
{Condon}, J. 2015, ArXiv e-prints

\bibitem[{{Doria} {et~al.}(2012){Doria}, {Gitti}, {Ettori}, {Brighenti},
  {Nulsen}, \& {McNamara}}]{Doria_2012}
{Doria}, A., {Gitti}, M., {Ettori}, S., {et~al.} 2012, \apj, 753, 47

\bibitem[{{Feretti} {et~al.}(2012){Feretti}, {Giovannini}, {Govoni}, \&
  {Murgia}}]{Feretti_2012}
{Feretti}, L., {Giovannini}, G., {Govoni}, F., \& {Murgia}, M. 2012, \aapr, 20,
  54

\bibitem[{{Gitti}(2013)}]{Gitti_2013b}
{Gitti}, M. 2013, \mnras, 436, L84

\bibitem[{{Gitti} {et~al.}(2012){Gitti}, {Brighenti}, \&
  {McNamara}}]{Gitti_2012}
{Gitti}, M., {Brighenti}, F., \& {McNamara}, B.~R. 2012, Advances in Astronomy,
  2012

\bibitem[{{Gitti} {et~al.}(2004){Gitti}, {Brunetti}, {Feretti}, \&
  {Setti}}]{Gitti_2004}
{Gitti}, M., {Brunetti}, G., {Feretti}, L., \& {Setti}, G. 2004, \aap, 417, 1

\bibitem[{{Gitti} {et~al.}(2002){Gitti}, {Brunetti}, \& {Setti}}]{Gitti_2002}
{Gitti}, M., {Brunetti}, G., \& {Setti}, G. 2002, \aap, 386, 456

\bibitem[{{Gitti} {et~al.}(2013){Gitti}, {Giroletti}, {Giovannini}, {Feretti},
  \& {Liuzzo}}]{Gitti_2013a}
{Gitti}, M., {Giroletti}, M., {Giovannini}, G., {Feretti}, L., \& {Liuzzo}, E.
  2013, \aap, 557, L14

\bibitem[{{Govoni} \& {Feretti}(2004)}]{Govoni-Feretti_2004}
{Govoni}, F. \& {Feretti}, L. 2004, International Journal of Modern Physics D,
  13, 1549

\bibitem[{{Kale} \& {Gitti}(2017)}]{Kale_2017}
{Kale}, R. \& {Gitti}, M. 2017, \mnras, 466, L19

\bibitem[{{Meisenheimer} {et~al.}(1997){Meisenheimer}, {Yates}, \&
  {Roeser}}]{Meisenheimer_1997}
{Meisenheimer}, K., {Yates}, M.~G., \& {Roeser}, H.-J. 1997, \aap, 325, 57

\bibitem[{{Mohr} {et~al.}(1996){Mohr}, {Geller}, \& {Wegner}}]{Mohr_1996}
{Mohr}, J.~J., {Geller}, M.~J., \& {Wegner}, G. 1996, \aj, 112, 1816

\bibitem[{{Murgia} {et~al.}(1999){Murgia}, {Fanti}, {Fanti}, {Gregorini},
  {Klein}, {Mack}, \& {Vigotti}}]{Murgia_1999}
{Murgia}, M., {Fanti}, C., {Fanti}, R., {et~al.} 1999, \aap, 345, 769

\bibitem[{{Pacholczyk}(1970)}]{Pacholczyk_1970}
{Pacholczyk}, A.~G. 1970, {Radio astrophysics. Nonthermal processes in galactic
  and extragalactic sources}, ed. {Pacholczyk, A.~G.}

\bibitem[{{Pizzolato} \& {Soker}(2005)}]{Pizzolato_2005}
{Pizzolato}, F. \& {Soker}, N. 2005, Advances in Space Research, 36, 762

\bibitem[{{Poggianti} {et~al.}(2016){Poggianti}, {Fasano}, {Omizzolo},
  {Gullieuszik}, {Bettoni}, {Moretti}, {Paccagnella}, {Jaff{\'e}}, {Vulcani},
  {Fritz}, {Couch}, \& {D'Onofrio}}]{Poggianti_2016}
{Poggianti}, B.~M., {Fasano}, G., {Omizzolo}, A., {et~al.} 2016, \aj, 151, 78

\bibitem[{{Rau} \& {Cornwell}(2011)}]{Rau_2011}
{Rau}, U. \& {Cornwell}, T.~J. 2011, \aap, 532, A71

\bibitem[{{Struble} \& {Rood}(1999)}]{Struble-Rood_1999}
{Struble}, M.~F. \& {Rood}, H.~J. 1999, \apjs, 125, 35

\bibitem[{{Wong} {et~al.}(2008){Wong}, {Sarazin}, {Blanton}, \&
  {Reiprich}}]{Wong_2008}
{Wong}, K.-W., {Sarazin}, C.~L., {Blanton}, E.~L., \& {Reiprich}, T.~H. 2008,
  \apj, 682, 155

\end{thebibliography}

\end{document}